\documentclass{emulateapj}

\newcommand{\Fig}[1]{Fig.~\ref{#1}}
\newcommand{\Sec}[1]{\S~\ref{#1}}
\newcommand{\Eqn}[1]{Eq.~\ref{#1}}
\newcommand{\lya}{Ly$\alpha$ }
\newcommand{\trec}{\ensuremath{t_{\rm rec}}}
\newcommand{\nbar}[1]{\ensuremath{\bar{n}_{\rm #1}}}
\newcommand{\pow}[2]{\ensuremath{#1 \times 10^{#2}}}
\newcommand{\hMpc}{\ensuremath{\,h^{-1}\,{\rm Mpc}}}
\newcommand{\ihMpc}{\ensuremath{\,h\,{\rm Mpc^{-1}}}}
\newcommand{\K}{\mbox{ K}}
\newcommand{\avg}[1]{\ensuremath{\langle #1 \rangle}}

\shorttitle{\lya Forest Power Spectrum}
\shortauthors{Lai et al.}

\begin{document}


\submitted{Accepted for publication in \apj}

\title{The Impact of Temperature Fluctuations on the \lya Forest Power
Spectrum}
\author{Kamson Lai\altaffilmark{1}, Adam Lidz\altaffilmark{1}, Lars
Hernquist\altaffilmark{1}, Matias Zaldarriaga\altaffilmark{1,2}}
\altaffiltext{1}{Harvard-Smithsonian Center for Astrophysics, 60
Garden Street, Cambridge, MA 02138, USA}
\altaffiltext{2}{Jefferson Laboratory of Physics, Harvard University,
Cambridge, MA 02138, USA}

\begin{abstract}
We explore the impact of spatial fluctuations in the intergalactic
medium temperature on the Ly$\alpha$ forest flux power spectrum near
$z \sim 3$. We develop a semianalytic model to examine temperature
fluctuations resulting from inhomogeneous \ion{H}{1} and incomplete
\ion{He}{2} reionizations. Detection of these fluctuations might
provide insight into the reionization histories of hydrogen and
helium. Furthermore, these fluctuations, neglected in previous
analyses, could bias constraints on cosmological parameters from the
Ly$\alpha$ forest.  We find that the temperature fluctuations
resulting from inhomogeneous \ion{H}{1} reionization are likely to be very
small, with an rms amplitude of $\la 5\%$, $\sigma_{T_0}/\langle T_0
\rangle \la 0.05$.  More important are the temperature fluctuations
that arise from incomplete \ion{He}{2} reionization, which might
plausibly be as large as $50\%$, $\sigma_{T_0}/ \langle T_0 \rangle
\sim 0.5$.  In practice, however, these temperature fluctuations have
only a small effect on flux power spectrum predictions.  The smallness
of the effect is possibly due to density fluctuations dominating over
temperature fluctuations on the scales probed by current measurements.
On the largest scales currently probed, $k \sim 0.001$ s km$^{-1}$
($\sim$0.1 $h$ Mpc$^{-1}$), the effect on the flux power spectrum may
be as large as $\sim 10\%$ in extreme models. The effect is larger on
small scales, up to $\sim 20\%$ at $k = 0.1$ s km$^{-1}$, due to thermal
broadening. Our results suggest that the omission of temperature
fluctuations effects from previous analyses does not significantly
bias constraints on cosmological parameters.
\end{abstract}


\section{Introduction} \label{intro}
In the current theoretical picture of the \lya forest, most of the
structure in the forest is a product of gravitational instability. The
absorbing gas is assumed to be in photoionization equilibrium with a
spatially homogeneous radiation field. On large scales the hydrogen
gas distribution follows the dark matter distribution, and on small
scales it is Jeans pressure-smoothed \citep[see e.g.,][]{cen94,
zhan95, hern96, mira96, muec96, bi97, bond97, hui97a, crof98, brya99,
dave99, theu99, nuss99}.  This gravitational instability model of the
forest, motivated by numerical simulations, seems to agree well with
observations \citep[e.g.][]{crof02, mcdo04b, tytl04, viel04c, lidz05}.

In this model, each \lya forest spectrum provides a one-dimensional
map of the density field in the intergalactic medium (IGM).  The \lya
forest can thus be used to constrain the amplitude and slope of the
linear matter power spectrum at $z \sim 3$ on scales of $k \sim 0.1 -
5 \ihMpc$ \citep[see e.g.,][]{crof98, mcdo00, zald01b, crof02,
zald03}.  When combined with measurements from Cosmic Microwave
Background (CMB) experiments and galaxy surveys, the \lya forest
provides important constraints on cosmological parameters
\citep[e.g.][]{selj05, viel05a}. Recently, very tight constraints on
cosmological parameters were derived using measurements of the \lya
forest flux power spectrum from the Sloan Digital Sky Survey (SDSS)
\citep{mcdo04a, mcdo04b, selj05}. The data samples used in the SDSS
flux power spectrum measurements are almost two orders of magnitude
larger than those used in previous measurements. The increase in
statistical precision sets a high bar for the required control over
systematic effects in theoretical predictions. The key issues are now
to devise consistency checks for the gravitational instability model
of the \lya forest, to quantify the accuracy of our theoretical
modeling, and to improve the modeling when possible.  These steps are
essential in order to estimate the systematic-error budget in the
modeling, and to utilize the full statistical power of the SDSS
measurements.

Toward this end, we point out that previous models of the \lya forest
adopt an over-simplified description of the IGM reionization history
and the resulting thermal evolution.  Specifically, previous models
assume that reionization is sudden and uniform, so that in effect each
gas element in the IGM experiences the same reionization history. In
reality, reionization is likely to be an extended and inhomogeneous
process \citep{soka03, soka04, bark04, furl04, babi05}, with some gas
elements reionizing earlier than others, and hence cooling to lower
temperatures by $z \sim 3$ \citep[e.g.][]{hui03}. Furthermore, \ion{He}{2}
may be reionized by bright quasars, and the process may be incomplete
near $z \sim 3$. In this case, the IGM at $z \sim 3$ resembles a
two-phase medium. The first phase consists of regions that have
already been engulfed by the \ion{He}{3} ionization fronts that are
expanding around bright quasars.  These regions, recently photo-heated
by a hard quasar spectrum, may have temperatures in excess of $\sim
\pow{3}{4}\K$ \citep{abel99}. The second phase consists of regions
where \ion{He}{2} has yet to reionize, but have \ion{H}{1}/\ion{He}{1} reionized at early
times. In this phase, gas elements with density near the cosmic mean
will be significantly cooler, with temperatures of $\sim
\pow{1}{4}\K$. These temperature fluctuations should be imprinted in
the \lya forest since the widths of \lya absorption lines, as well as
the hydrogen recombination coefficient, and hence the optical depth to
\lya absorption, depend on temperature.

The goal of this paper is to estimate the amplitude and spatial scale
of the temperature fluctuations resulting from \ion{H}{1} and \ion{He}{2}
reionization, and to examine their impact on the flux power spectrum.
Temperature fluctuations are particularly interesting because if they
significantly impact the flux power spectrum, then their detection
would likely provide insights into the reionization histories of
hydrogen and helium. Furthermore, we reiterate that it is important to
check whether omitting temperature fluctuations in the analyses will
significantly bias the determination of cosmological parameters from
the \lya forest.

The outline of this paper is as follows. We begin with a brief
overview of the theoretical model describing the \lya forest in
\Sec{review}. We then estimate the amplitude of temperature
fluctuations expected from inhomogeneous \ion{H}{1} reionization in
\Sec{tfluc_hi}.  In \Sec{tfluc_heii} we estimate the amplitude and
scale of temperature fluctuations for a range of models describing
\ion{He}{2} reionization by bright quasars, using the observed quasar
luminosity function as input. In \Sec{flux_power}, we examine the
impact of these fluctuations on the flux power spectrum.  We conclude
in \Sec{conclusion} and discuss possible future research directions.

\section{Modeling the \lya Forest} \label{review}

In this section, we briefly review the standard theoretical model of
the \lya forest in order to introduce notation, and highlight the
approximations that we subsequently test in this paper.  For more
details, the reader can refer to e.g., \citet{hui97a}.

The gas responsible for the absorption in the \lya forest is thought
to be in photoionization equilibrium with a
radiation background produced by star-forming galaxies and/or
quasars. In this case, the abundance of neutral hydrogen scales like
$n_{\rm HI} \propto \alpha(T) \Delta^2/\Gamma$. Here, $\alpha(T)
\propto T^{-0.7}$ is the temperature dependent hydrogen recombination
coefficient, $\Delta$ is the baryon density in units of the cosmic
mean, and $\Gamma$ is the hydrogen photoionization rate.  The optical
depth to \lya absorption is proportional to the neutral hydrogen
abundance (besides thermal broadening and peculiar velocities),
which implies a simple power-law relationship between the \lya optical
depth and the gas density.  Therefore, if the gas temperature and
$\Gamma$ are known or can be modeled, then there exists a direct
connection between the \lya absorption and the underlying density
fluctuations.

Indeed, the physics that sets the temperature of the absorbing gas is
expected to be relatively simple. The gas temperature is determined
largely by the competition between photoionization heating and
adiabatic cooling \citep{mira94, hui97b}. In this case, the
temperature of the low density gas, where shock-heating should be
unimportant, is expected to be tightly correlated with its density
\citep{hui97b}. In fact, these authors show that the gas temperature
should be a power-law in the gas density: $T = T_0 \Delta^{\gamma
-1}$.  The numerical values of the power law index, $\gamma$, and the
temperature at mean density, $T_0$, depend on when the gas was
reionized and the nature of the ionizing sources
\citep[e.g.][]{hui97b, abel99, soka02, theu02a, hui03}.

The standard assumption is that $\Gamma$, $T_0$ and $\gamma$ are all
spatially uniform, i.e. they have a single value throughout the
entire IGM. To the extent that this is true, the neutral hydrogen
density, and hence the \lya optical depth, scales as $\tau = A_\tau
\Delta^{2 - 0.7(\gamma -1)}$, where the proportionality constant
$A_\tau$ is independent of spatial position.  In this simple case,
fluctuations in the \lya forest absorption directly traces the
underlying density fluctuations.  Several authors have investigated
the possible `contamination' from spatial fluctuations in the hydrogen
photoionization rate, $\Gamma$, finding that these fluctuations should
be quite small near $z \sim 3$ \citep{crof99, meik04, crof04,
mcdo04c}.  Little attention, however, has been given to the assumption
that $T_0$ and $\gamma$ are also spatially uniform.

In the case where $T_0$ and $\gamma$ fluctuate spatially, the relation
for the \lya optical depth generalizes to:
\begin{equation} \label{Eqn::tau}
\tau = A_\tau \Delta^{2 - 0.7(\gamma-1)} (1+\delta_{T_0})^{-0.7}.
\end{equation}
The new ingredient in this equation is the term $1+\delta_{T_0} = T_0
/ \avg{T_0}$ which represents spatial fluctuations in the temperature
of the gas at the cosmic mean density. Furthermore, we consider
spatial fluctuations in $\gamma$, i.e.\ $\gamma$ in this equation now
depends on position.  Additionally, fluctuations in the IGM
temperature lead to spatial variations in the thermal broadening
kernel, which will affect the \lya forest on small scales.

The flux transmitted through the \lya forest, after taking into
account peculiar velocities and thermal broadening, is given by $F =
e^{-\tau}$.  Fluctuations in the transmission are given by $\delta_F =
(F - \avg{F})/\avg{F}$, and the power spectrum of $\delta_F$ is the
flux power spectrum, which we denote by $P_F(k)$.  Our goal then is to
explore the impact of the temperature fluctuations encoded in
\Eqn{Eqn::tau}, and in the thermal broadening kernel, on the flux
power spectrum $P_F(k)$.

\section{Temperature Fluctuations from Inhomogeneous \ion{H}{1} Reionization}
\label{tfluc_hi}

\begin{figure}[t]
\includegraphics[width=\columnwidth,bb=55 330 470 740]{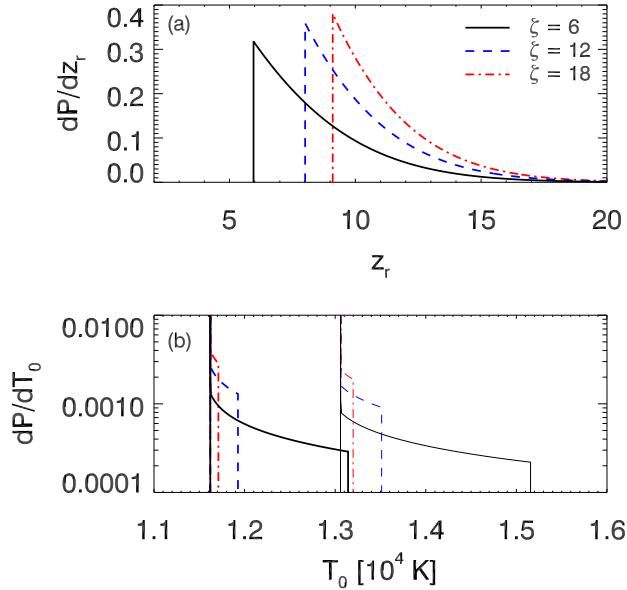}
\caption{{\it Panel (a):} \ion{H}{1} reionization redshift distribution
for $\zeta$ = 6, 12, and 18.  {\it Panel (b):} Temperature distribution
at $z=3$ for the same reionization models in (a).  The thick (thin)
curves are calculated using $T_r = 2.5 \times 10^4$ K ($3 \times 10^4$
K).}
\label{Fig::HI}
\end{figure}

We begin by considering the amplitude of temperature fluctuations
resulting from \ion{H}{1} reionization, which recent theoretical work has
emphasized to be likely inhomogeneous and extended \citep{soka03,
bark04, furl04, babi05}.  We use the model of \citet{furl04} to
describe the duration of extended \ion{H}{1} reionization, and scaling
relations from \citet{hui97b} to estimate the amplitude of the
resulting temperature fluctuations.  Throughout this paper, we will
work generally within the context of a model in which \ion{H}{1}/\ion{He}{1} are
reionized at high redshift by star-forming galaxies, while \ion{He}{2} is
reionized close to $z \sim 3$ by quasars.  This is not the only
possibility (see Lidz et al. 2005, in prep., for a discussion), and so
it is useful for us to explicitly separate out the temperature
fluctuations that arise from \ion{H}{1} reionization and those that arise from
\ion{He}{2} reionization.  The case that \ion{H}{1} and \ion{He}{2} are both reionized at
high redshift will then be similar to our \ion{H}{1} calculation: the only
difference being that the temperature at reionization would likely be
higher, and the duration of the reionization process might be
different.

The \citet{furl04} model describes the growth and overlap of \ion{H}{2}
regions during reionization. The basic picture in this model is that
large scale overdense regions contain more ionizing sources, and are
reionized earlier, than underdense regions.  The model assumes that a
galaxy of mass $m_{\rm gal}$ can ionize a mass corresponding to $\zeta
m_{\rm gal}$, where $\zeta$ is an unknown parameter describing how
efficiently a galaxy can ionize surrounding gas (see \citealt{furl05b}
and \citealt{furl05a} for extensions to this model).  A region is
considered ionized when the the fraction of mass in halos more massive
than some minimum mass, $m_{\rm min}$, exceeds a threshold set by the
ionization efficiency of the sources: $f_{\rm coll} >
\zeta^{-1}$. Here, $f_{\rm coll}$ denotes the fraction of mass in the
region which has collapsed into halos of mass larger than $m_{\rm
min}$.  In this case, the size distribution of \ion{H}{2} regions can be
calculated in a similar manner to the halo mass function in the
excursion set formalism \citep[e.g.,][]{bond91, lace93}.

The \citet{furl04} model predicts, given $\zeta$, the filling factor
of ionized regions.  It is given by
\begin{equation} \label{Eqn::Q_HI}
  Q = \zeta \bar{f}_{\rm coll}
\end{equation}
where $\bar{f}_{\rm coll}$ is the global collapse fraction (different from
$f_{\rm coll}$, the collapse fraction of a region with a given
overdensity).  In the extended Press-Schechter theory, the global
collapse fraction is given by \citep{bond91, lace93}:
\begin{equation} \label{Eqn::f_coll}
  \bar{f}_{\rm coll} = {\rm erfc} \left[ \frac{\delta_{\rm c}(z)}{\sqrt{2
    \sigma_{\rm min}^2(z)}} \right],
\end{equation}
where $\delta_{\rm c}(z)$ is the critical density for collapse and
$\sigma_{\rm min}^2(z)$ is the density variance on the scale of
$m_{\rm min}$, the minimum mass of an ionizing source. We take the
minimum mass to be the mass corresponding to a virial temperature of
$10^4 \K$, where atomic hydrogen line cooling is efficient.

The probability distribution of reionization redshift is related to
the filling factor $Q$ by
\begin{equation} \label{Eqn::Pz}
  \frac{dP}{dz_r} = \left. -\frac{dQ}{dz} \right|_{z = z_r}.
\end{equation}
In \Fig{Fig::HI}a, we plot $dP/dz_r$ for $\zeta$ = 6, 12, and 18.  The
curves terminate when $Q = 1$, corresponding to the end of
reionization.  The figure clearly illustrates that \ion{H}{1} reionization
should be quite extended, with the whole process taking place over a
$\Delta z$ of several.  The end, and somewhat the duration, of
reionization naturally depends on how efficiently the sources produce
ionizing photons, which is quite uncertain. We therefore consider
$\zeta$ = 6, 12, and 18, in which case the end of reionization occurs
at $z =$ 6, 8, and 9 respectively.  For our purposes, these choices of
$\zeta$ bracket the interesting range of possibilities.  It is
uninteresting to consider sources that are much less efficient,
because we know the IGM is highly ionized below $z \sim 6$
\citep[e.g.][]{fan02}. On the other hand, more efficient sources would
lead to a very early end to reionization, in which case the
temperature at lower redshifts is completely insensitive to when
precisely the gas was reionized owing to efficient Compton cooling
\citep{hui03}.

In order to investigate the temperature fluctuations that result from
extended \ion{H}{1} reionization, we rely on analytic approximations by
\citet{hui97b} to describe the thermal history of the IGM. These
analytic approximations derive from the fact that the temperature of
the low density gas is primarily determined by photoionization heating
and adiabatic cooling. Under these simple physical conditions,
\citet{hui97b} give formulae for the evolution of $T_0$ and $\gamma$
(see their Eq.~19 and 22). The input to these formulae is simply the
temperature, $T_r$, that a gas element reaches when it is reionized at
redshift $z_r$. Given $T_r$, these formulae give the values of $T_0$
and $\gamma$ at lower redshifts, $z < z_r$.

The temperature at reionization, $T_r$, is quite uncertain. It is
determined both by the intrinsic spectrum of the ionizing sources, and
the hardening of the spectrum owing to absorption in the IGM
\citep{abel99}. Scatter in the intrinsic and re-processed spectra will
likely give rise to a distribution of $T_r$. It is also unclear that
gas elements reionized at different times, albeit by sources with the
same intrinsic spectrum, will reach the same temperature following
reionization.  We currently ignore these subtleties in our model, and
assume that all gas elements will reionize to the same $T_r$.  The
value $T_r = \pow{2.5}{4}\K$ is reasonable assuming that galaxies are
the ionizing sources \citep{hui97b}.  We also investigate the effects
of using a more extreme $T_r = \pow{3}{4}\K$.  Note that because of
the high abundance of galaxies, each gas element will likely see the
combined radiation from numerous sources during reionization.  This
will tend to average out the scatter in the intrinsic and re-processed
spectra, so a uniform $T_r$ is probably a good approximation.  The
same may not be true if quasars are the ionizing sources, since
quasars are sparse and each gas element will only see the radiation
from one, or a few, nearby sources.

With $z$ and $T_r$ fixed, $T_0$ is a function of $z_r$ only and we can
find the distribution of $T_0$ with the simple transformation:
\begin{equation} \label{Eqn::Pt}
  \frac{dP}{dT_0} = \frac{dP}{dz_r} \left|\frac{dT_0}{dz_r}\right|^{-1}.
\end{equation}
The temperature distribution is plotted in \Fig{Fig::HI}b.  The cutoff
in the distribution at high $T_0$ occurs because there is a definite
end to reionization in our model, after which the probability of
reionization is formally zero.  The upper temperature limit to the
distribution is therefore set by the $T_0$ of the most recently
reionized gas elements.  There is a sharp rise in the temperature
distribution at low $T_0$, because gas elements reionized above $z
\sim 10$ will have reached almost the same temperature at low
redshift, owing to efficient Compton cooling at high redshift
\citep{hui97b}.

The temperature distribution at $z=3$ spans a very narrow range in
$T_0$, generally less than a few hundred degrees Kelvin, in all the
models we consider. This is because the interplay between
photoionization heating and adiabatic cooling drives the gas towards a
thermal asymptote \citep{hui03}.  Therefore, gas elements that are
reionized sufficiently early will approach the same asymptotic
temperature at low redshifts.  As a result, the temperature
fluctuations at $z=3$ owing to \ion{H}{1} reionization are small.  In the
$\zeta=6$ model, with $T_r = 2.5 \times 10^4$ K, the level of
temperature fluctuations is about $4\%$, $\sigma_{T_0}/\avg{T_0} =
0.04$.  In the $\zeta = 12$ and 18 models, the levels of temperature
fluctuations are even smaller at $< 1\%$.  Taking $T_r = 3 \times
10^4$ K does not give significantly larger temperature fluctuations.
The amplitude of temperature fluctuations might be larger at higher
redshifts, because the gas elements have less time to cool.  However,
efficient cooling quickly causes the gas to approach the thermal
asymptote, and we find that the amplitude of temperature fluctuations
at $z=4$ is very similar to that at $z=3$.

It is possible to have temperature fluctuations larger than what we
have estimated.  For instance, if \ion{He}{2} is reionized alongside \ion{H}{1}/\ion{He}{1},
then temperatures of $\pow{4}{4}\K$ or higher are plausible following
reionization \citep{abel99}, leading to larger temperature
fluctuations.  Furthermore, the analytic approximation we employ
assumes a constant spectrum for the ionizing background.  In reality,
the spectrum experienced by a gas element can be hardened
significantly during reionization relative to the late time spectrum
\citep{abel99}.  Therefore, for a given $T_r$, the analytic
approximation tends to overestimate the late time temperature, since
the hardened spectrum at reionization is assumed throughout.
Processes such as recombination cooling are also neglected in the
analytic approximation, further contributing to the overestimation of
the late time temperature.  A more detailed calculation taking into
account different cooling mechanisms and variations in the spectrum of
the ionizing background will in general yield larger temperature
fluctuations.  Note also that the temperature fluctuations at $z \sim
6$ might be significantly larger than those at $z \sim 3$.  This might
bias constraints derived from $z \sim 6$ quasar spectra, such as the
evolution of the hydrogen photoionization rate \citep[e.g.][]{fan02}.

In summary, we find that while \ion{H}{1} reionization can be quite extended,
gas cooling erases temperature fluctuations over time.  Therefore,
temperature fluctuations from inhomogeneous \ion{H}{1} reionization are likely
negligible at $z=3$, especially if \ion{H}{1} is reionized by $z \ga 8$.  Even
though possibilities such as a high temperature at reionization or
evolution in the spectrum of the ionizing background might lead to
larger temperature fluctuations, it seems likely that there must be
`reionization activity' very near $z \sim 3$ in order for the
temperature to fluctuate significantly at this redshift.

\section{Temperature Fluctuations from Incomplete \ion{He}{2} Reionization}
\label{tfluc_heii}

\begin{deluxetable*}{lccccccc}
\tablewidth{0pt}
\tablecaption{Model Parameters \label{ParTab}}
\tablehead{
  \colhead{Model}           & \colhead{$\alpha(z=2.1)$}     &
  \colhead{$\alpha(z>3.6)$} & \colhead{$\beta$}             &
  \colhead{$A$}             & \colhead{$B$}                 &
  \colhead{$t_q$ [yrs]}       & \colhead{$T_r$ [K]}}
\startdata
Fiducial         & -3.28 & -2.58 & -1.78 & -7.31 & 0.47 & $10^7$ & \pow{3}{4} \\
Small Fluctuations (SF) & -3.48 & -2.81 & -1.78 & -7.12 & 0.32 & $10^7$ & \pow{3}{4} \\
Large Fluctuations (LF) & -3.28 & -2.58 & -1.78 & -7.69 & 0.77 & $10^7$ & \pow{4}{4} \\
Small Scale (SS) & -3.28 & -2.58 & -1.78 & -7.69 & 0.77 & $10^6$ & \pow{4}{4} \\
Large Scale (LS) & -3.28 & -2.58 & -1.78 & -7.50 & 0.77 & $10^8$ & \pow{4}{4} \\
Large Fluc. $z=4$ (LF4) & -3.28 & -2.58 & -1.78 & -7.12 & 0.47 &
$10^7$ & \pow{4}{4} 
\enddata
\tablecomments{\small Parameters for the QLF ($\alpha$, $\beta$, $A$,
and $B$) are based on the best fit values and uncertainties in
\citet{rich05} and \citet{fan01a}.  The normalization of the QLF from
\citet{rich05}, $\phi_* = \pow{5.96}{-7} {\rm Mpc^{-3} mag^{-1}}$, is
used in all models.}
\end{deluxetable*}

One possible scenario that can give rise to potentially large
temperature fluctuations is when \ion{He}{2} is reionized gradually by bright
quasars.  If \ion{He}{2} reionization is still underway at $z \sim 3$, there
should be hot \ion{He}{3} bubbles around bright quasars embedded in a much
cooler background IGM, in which only \ion{H}{1}/\ion{He}{1} has reionized. The
temperature in the hot bubbles may be as high as $T_0 \sim \pow{3}{4}$
K \citep{abel99}, while the temperature outside of these bubbles
should be around $T_0 \sim 10^4$ K, as described in the previous
section. In this case, the temperature fluctuations can be
significantly larger than those arising from \ion{H}{1} reionization.
Additionally, the regions that recently reionized \ion{He}{2} will be close
to isothermal ($\gamma = 1$), while the cool exterior will have a
steeper temperature-density relation, $\gamma \sim 1.4 - 1.6$, with
the precise value depending on when \ion{H}{1}/\ion{He}{1} reionizes.  In this
section, we begin by describing a simple model for the growth of \ion{He}{3}
bubbles, and their subsequent thermal evolution.  We will then discuss
the resulting temperature distribution and power spectrum of
temperature fluctuations.

\subsection{Numerical Model} \label{model}
Our procedure is to populate a simulation box with quasars by drawing
sources from the observed quasar luminosity function (QLF).  We then
follow the growth of \ion{He}{3} bubbles around each quasar source for a
fixed quasar lifetime, $t_q$, and record the subsequent thermal
evolution inside the ionized regions.  We can then measure the
statistics of the resulting temperature field, and use the temperature
field, along with the density and peculiar velocity fields from a
cosmological simulation, to study the statistics of the \lya forest.

First, we will consider the time evolution of the \ion{He}{3} ionized volume
around an isolated quasar source.  The most interesting period for
temperature fluctuations is before the overlap of \ion{He}{3} ionized
regions is complete.  In this pre-overlap phase, the growth of \ion{He}{3}
regions is simple to describe, since we can make the approximation
that a gas element will only see the radiation from the central
quasar.  The growth of the ionized region around the quasar is then
given by \citep{shap87, mada99}:
\begin{equation} \label{Eqn::dVdt}
\frac{dV}{dt} = \frac{\dot{N}}{\nbar{He}} - \frac{V}{\trec}.
\end{equation}
In the above equation, $V$ is the {\em comoving volume} of the ionized
bubble, \nbar{He} is the cosmic mean comoving number density of helium
atoms, and $\dot{N}$ is the number of \ion{He}{2} ionizing photons emitted by
the quasar per unit time. This equation assumes that all of the helium
in the pre-\ion{He}{2} reionized gas is singly ionized.  Further, it assumes
that all photons emitted above the \ion{He}{2} threshold contribute to \ion{He}{2}
reionization, since absorption of \ion{He}{2} ionizing photons by \ion{H}{1} and \ion{He}{1}
is negligible owing to their high ionization levels.  The ionized
region is assumed to be spherical, and grows according to
\Eqn{Eqn::dVdt} while the quasar is active.  The bubble is assumed to
remain fixed at its final size after the quasar turns off, and the
subsequent redshift evolution of $T_0$ and $\gamma$ inside the bubble
is tracked by the formulae of \citet{hui97b}.

In \Eqn{Eqn::dVdt}, a single volume averaged recombination rate is
assumed for \ion{He}{3} inside the bubble.  The recombination time is $\trec
= \left( \nbar{e,p} \alpha^B_{\rm HeIII} C \right)^{-1}$, where
$\alpha^B_{\rm HeIII}$ is the recombination coefficient to the excited
states of \ion{He}{3} (case B) \citep[see][]{mada99}, and \nbar{e,p} is the
{\em proper} electron number density.  The clumping factor of \ion{He}{3} is
defined as $C = \avg{n^2_{\rm HeIII}} / \nbar{HeIII}^2$.  The value of
$C$ is quite uncertain, and values between $C = 1 - 30$ are commonly
used in the literature \citep[see e.g.,][]{mada99, meik05}.  In our
calculation we chose $C = 1$. Regardless, the effect of recombinations
on the growth of \ion{He}{3} bubbles should be limited, since the
recombination time \trec\ is much longer than the expected quasar
lifetime $t_q$.  Assuming an IGM temperature of \pow{3}{4}\K, \trec\
is on the order of $10^9$ yrs at $z=3$, much longer than $t_q$, which
we vary between $10^6 - 10^8$ yrs.  A quasar of luminosity $L$ will
then be surrounded by a \ion{He}{3} region with an approximate volume of $V
\propto L t_q$ at the end of its lifetime.

The next ingredient in our modeling is our description of the
abundance, spectrum, and lifetime of quasars. We parameterize the QLF
with the standard double power law \citep{boyl98, pei95, croo04}:
\begin{equation} \label{Eqn::QLF}
  \phi(L,z) = \frac{\phi_*/L_*}{(L/L_*)^{-\alpha}+(L/L_*)^{-\beta}}.
\end{equation}
We use measurements of the QLF from 2SLAQ \citep{rich05} at $0.4 < z <
2.1$, and SDSS \citep{fan01a} at $z > 3.6$.  Note that the bright-end
slope $\alpha$ from SDSS appears to be significantly different
($\gtrsim 2 \sigma$) than that measured by 2SLAQ at low redshift.
Also, the SDSS measurements at $z > 3.6$ do not probe the faint end of
the QLF.  In order to bridge the gap between the SDSS and 2SLAQ
measurements, and to extrapolate to all relevant luminosities, we make
several assumptions.  First, at intermediate redshifts, we linearly
interpolate between the low and high redshift bright-end
slopes. Second, we use the faint-end slope, $\beta$, and the
normalization of the QLF, $\phi_*$, from 2SLAQ and assume that they
remain fixed with redshift.  Finally, we fixed $L_*$ by requiring that
our model QLF matches the SDSS best-fit abundance of bright
quasars. Specifically, we match to the fitting formula of
\citet{fan01a}\footnotemark, in which the number density of quasars
with magnitudes $M_{1450} < -26$ is given by $\log \Phi(z,M_{1450} <
-26) = A - B(z-3)$, with $\Phi$ in units of Mpc$^{-3}$.

\footnotetext{The cosmology adopted by Fan et al.\ (\{$\Omega_{\rm
m}$, $\Omega_{\rm \Lambda}$, $h$\} = \{0.35, 0.65, 0.65\}) is slightly
different from the one we use (\{$\Omega_{\rm m}$, $\Omega_{\rm
\Lambda}$, $h$\} = \{0.3, 0.7, 0.7\}), but we do not attempt to
correct for this small difference.}

Finally, we adopt a low luminosity cutoff for the quasar luminosity
function of $L_{\rm min} = 0.018 L_*$ (see \citealt{mada99} for a
discussion), and the quasar spectrum from \citet{mada99}, which goes
as $\nu^{-1.8}$ above the \ion{He}{2} ionization threshold. All quasar
sources are assumed to have the same spectrum and lifetime.

With these considerations in mind, we investigate several models. Our
strategy here is to span a conservative range in the parameters that
characterize the quasar sources, and hence a range in the amplitude
and characteristic scale of the resulting temperature
fluctuations. This is prudent given not only the observational
uncertainties in these parameters, but also the uncertainties and
approximations inherent in our modeling. The parameters of these
models are summarized in Table~\ref{ParTab}. In our fiducial model, we
adopt the best fit values from 2SLAQ and SDSS for the parameters in
the QLF, and use $t_q = 10^7$ yrs and $T_r = \pow{3}{4}\K$ for the
quasar lifetime and temperature at reionization, respectively.  We
then vary the parameters around these values, investigating models
with small temperature fluctuations (SF), and large temperature
fluctuations (LF), as detailed in Table~\ref{ParTab}. In the LF model,
in addition to varying the parameters of the QLF, we adopt a large
temperature at reionization, $T_r = \pow{4}{4}\K$. In each case, we
vary the parameters of the observed QLF within their allowed
2-$\sigma$ range.  Finally, we vary the quasar lifetime in order to
cover a range in the characteristic scale of the temperature
fluctuations.  Specifically, the large scale (LS) temperature
fluctuations model adopts $t_q = 10^8$ yrs, while the small scale (SS)
temperature fluctuations model adopts $t_q = 10^6$ yrs.

\subsection{Temperature Distributions} \label{tdist}

\begin{figure}[t]
\includegraphics[width=\columnwidth,bb=85 330 460 740]{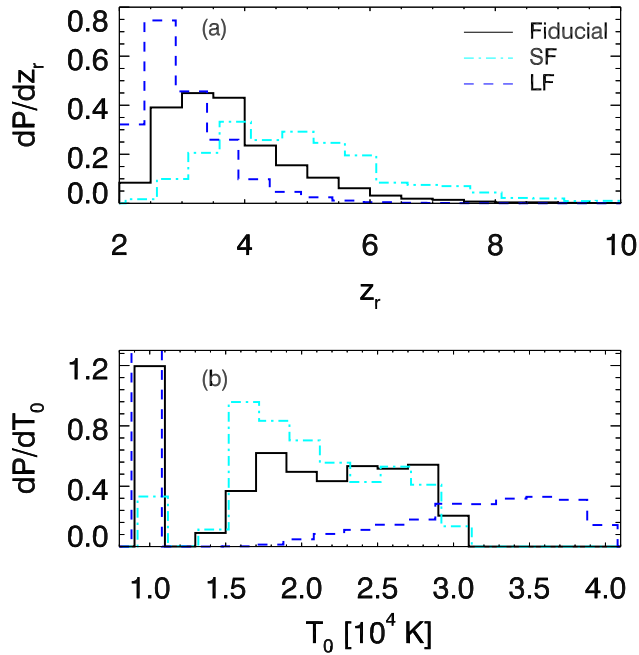}
\caption{{\it Panel (a):} \ion{He}{2} reionization redshift
distribution for the fiducial, SF, and LF models.  {\it Panel (b):}
Temperature distribution at $z=3$ for the same reionization models in
(a).  For illustrative clarity, the SF and LF models are shifted in
the $z_r$-direction by +0.1 and -0.1 respectively in (a), and in the
$T_0$-direction by +200 K and -200 K in (b).}
\label{Fig::HeII}
\end{figure}

We construct realizations of the $T_0$ and $\gamma$ fields in each of
the models discussed in the previous section.  We use a $80 \hMpc$
comoving box, and $512^3$ mesh points. This boxsize is convenient for
overlaying on the cosmological simulation which we will describe in
\Sec{flux_power}. Throughout, we average the statistics in the
fiducial, LF, and LS models over five independent realizations, in
order to reduce scatter owing to the small number of \ion{He}{3} bubbles in
our $80 \hMpc$ simulation box. In the other models, the bubble
distribution is sampled sufficiently well with a single realization.

With the numerical models in hand, we proceed to measure the
probability distribution of the temperature field. We first examine in
\Fig{Fig::HeII}a the probability distribution of \ion{He}{2} reionization
redshifts, in analogy with \Fig{Fig::HI}a. This figure is
constructed by recording the redshift at which each pixel in the
simulation is first engulfed by an expanding \ion{He}{3} bubble. We only
plot our fiducial, SF, and LF models, since the other models have
temperature distributions that are similar to that of the LF
model. One can see that \ion{He}{2} reionization is quite extended, and that
these simple models are each consistent with incomplete \ion{He}{2}
reionization near $z \sim 3$. Our models are essentially Monte-Carlo
versions of the \citet{mada99} calculation and are consistent with
these earlier calculations.
 
The resulting temperature distributions at $z=3$ are shown in
\Fig{Fig::HeII}b. In our models, we assume that the background IGM, in
which only \ion{H}{1}/\ion{He}{1} is reionized, has a uniform temperature at $T_0 =
10^4\K$, as justified in \Sec{tfluc_hi}. We additionally assume a
uniform $\gamma = 1.4$ for the background IGM. In the figure, one can
clearly see the bimodal temperature distribution that results from
incomplete \ion{He}{2} reionization.  The $\gamma$ distribution exhibits a
similar bimodal distribution, with $\gamma \sim 1$ inside \ion{He}{3}
regions, and $\gamma = 1.4$ in the background IGM. The temperature
distributions are quite broad, with r.m.s.\ fluctuation amplitudes of
$\sigma_{T_0} / \avg{T_0}$ = 0.24, 0.33, and 0.58 ($\sigma_{T_0}$ =
\pow{4.8}{3}\K, \pow{6.4}{3}\K, and \pow{1.2}{4}\K) in the SF,
fiducial, and LF models, respectively.  Note that the higher
temperatures of the hot regions in the LF model are not a consequence
of that model's reionization history, but rather because we chose a
larger $T_r$ to maximize fluctuations.  Compared to the results from
\Sec{tfluc_hi}, we see that temperature fluctuations from incomplete
\ion{He}{2} reionization can be as much as a factor of 10 larger than that
from extended \ion{H}{1} reionization.

The temperature distributions shown in \Fig{Fig::HeII}b are each
distributions at $z = 3$ in different models, but they also roughly
represent the fluctuations during different {\em stages} of
reionization. The filling factor of \ion{He}{3} regions, $Q$, is indicated
by the area under the hot component of the temperature
distribution. In the LF model, $Q$ is about 0.5 at $z = 3$. Here, the
amplitude of temperature fluctuations is maximal, as the hot and cold
regions occupy comparable fractions of space. On the other hand, in
the SF and fiducial models, reionization is considerably more
complete.  The temperature distribution is dominated by hot regions by
$z=3$, and the fluctuations are smaller.  At a higher redshift, \ion{He}{2}
reionization is less complete in the fiducial and SF models, and the
temperature distributions in these models will be similar to that in
the LF model at $z = 3$.

\subsection{Power Spectrum of Temperature Fluctuations} \label{Sec::TempPS}

\begin{figure*}[t]
\centering
\includegraphics[width=.9\textwidth,bb=40 380 625 750]{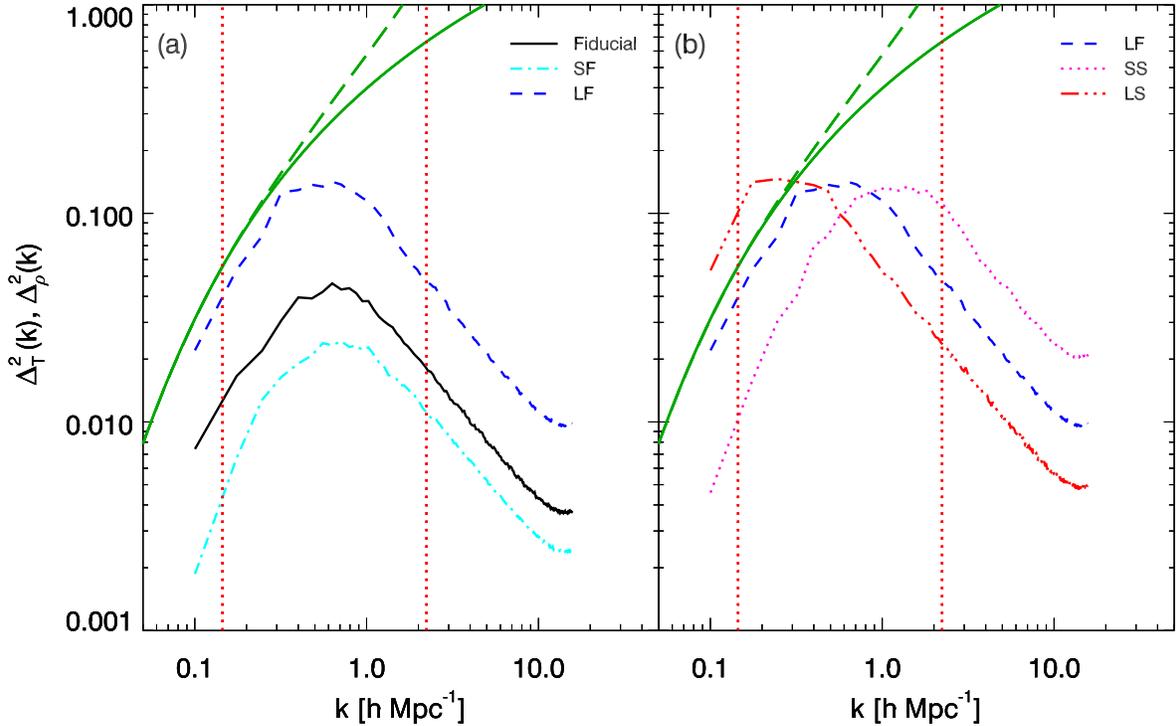}
\caption{{\it Panel (a):} 3-dimensional power spectrum of $T_0$ at
$z=3$ in the fiducial, LF, and SF models.  {\it Panel (b):}
3-dimensional power spectrum of $T_0$ at $z=3$ in the SS and LS
models, with the LF model repeated for comparison.  The linear and
non-linear matter power spectrum are shown as the green solid and
dashed lines, respectively.  The vertical red dotted lines indicate
the range of scales probed by SDSS.
\label{Fig::TPS}}
\end{figure*}

The temperature distributions presented above are informative, but
they tell us nothing about the {\em characteristic scale} of the
temperature fluctuations. To investigate this, we measure the
3-dimensional power spectrum of the temperature field, $T_0$, from our
simulations. In \Fig{Fig::TPS} we show the dimensionless power
spectrum, $\Delta_T^2 (k) \propto k^3 P_T(k)$, for each of the five
models we consider. Each power spectrum shows a well defined
characteristic scale, with the power spectrum growing as $k^3$ on
large scales, and falling as $k^{-1}$ on small scales.  The large and
small scales trends in $\Delta_T^2(k)$ are a natural consequence of
our model in which the source positions are uncorrelated (see below
for comments on this approximation).  The assumptions that the bubbles
have a well defined boundary, and that $T_0$ within each bubble is
uniform, also contribute to the trends seen in $\Delta_T^2(k)$. The
characteristic bubble size is set by (see Eqs.~\ref{Eqn::dVdt} and
\ref{Eqn::QLF}) the break in the luminosity function, $L_*$, the
quasar spectrum, and the quasar lifetime, $t_q$. The characteristic
comoving scale in the fiducial, SF, and LF models is $R_* \sim 5
\hMpc$, while it is $R_* \sim 11 \hMpc$ in the LS model, and $R_* \sim
2 \hMpc$ in the SS model.

Two simplifying assumptions in our model may cause the characteristic
scale to be underestimated.  First, our model does not take into
account quasar clustering, and so we tend to underestimate the chance
that ionized bubbles around neighboring quasars will overlap to form
large ionized regions.  Assuming a quasar bias $b_q = 4$ (extrapolated
from \citealt{croo05}), we find that clustering enhances the average
number of sources inside an ionized region by a factor of $\sim 3$
over that of a uniform distribution.  However, even when the effect of
clustering is included, there are on average $< 1$ additional active
sources inside an ionized region of size $R \sim 10 \hMpc$.
Therefore, the clustering of active sources can safely be ignored.
Quasar clustering will also lead to the clustering and overlap of
fossil ionized regions.  In this case, neglecting clustering may cause
the typical volume of hot regions to be underestimated by as much as a
factor of 3.  It is therefore prudent to investigate models spanning a
wide range of characteristic scales, as we have done.  The second
simplification in our model is in our treatment of multiple ionizing
sources.  When the ionization fronts of two or more bubbles overlap,
the resulting large ionized region will expand so as to conserve
ionizing photons coming from the multiple sources inside the combined
region.  We neglect this subtlety in our modeling, noting that the
probability for overlap is not significant around $Q = 0.5$, when
temperature fluctuations are largest.

We aim to explore the effects of temperature fluctuations on the
statistics of the absorption in the \lya forest.  Since in the
standard picture of the forest most structure derives from
gravitational instability, it is instructive to compare temperature
fluctuations with density fluctuations. Which is a more important
source of structure in the \lya forest, temperature or density
fluctuations? We address this question in \Fig{Fig::TPS}, where we
include curves indicating the linear matter power spectrum, and the
\citet{peac96} fit for the non-linear power spectrum. This may not be
exactly the relevant comparison: for instance, the optical depth to
\lya absorption scales with matter density as $(1+\delta)^2$, while it
only varies with temperature as $(1+\delta_{T_0})^{-0.7}$. Density
fluctuations of a given amplitude are therefore amplified into larger
optical depth fluctuations than temperature fluctuations of the same
amplitude. Nonetheless, the comparison is suggestive. \Fig{Fig::TPS}
illustrates that on small scales, $k \ga 1 \ihMpc$, the density power
is at least $1-2$ orders of magnitude larger than the temperature
power in all models considered. On these scales, fluctuations in the
density field will be the more important source of structure in the
\lya forest.  On the other hand, on large scales, the temperature
power may be comparable, or even dominant, in comparison to the
density power. For instance, in the LS model, the temperature power is
actually larger than the density power for $k \la 0.3 \ihMpc$.

Our results seem to suggest that temperature fluctuations may be an
important source of structure in the \lya forest on large
scales. There are, however, a few caveats to this intuition. First, as
mentioned above, the optical depth scales more strongly with density
than temperature. Second, current flux power spectrum measurements do
not probe very large scales, as illustrated by the red dotted lines in
\Fig{Fig::TPS}, which show the range of scales probed by SDSS
observations. This suggests that if the characteristic scale of the
temperature fluctuations is large, much of the effect will be on
scales larger than that probed by current measurements. Finally, it is
important to keep in mind, that the \lya forest provides a {\em 1-d
skewer} through the IGM.  As a result, the power spectrum of
fluctuations along a line of sight, on large scales, includes aliased
power from small wavelength modes transverse to the line of sight.
This may tend to wash out some of the signal from temperature
fluctuations on large scales.

\section{Impact of Temperature Fluctuations on the Flux Power Spectrum}
\label{flux_power}

We now arrive at the heart of our study: how do the temperature
fluctuations, described in the previous section, impact the flux power
spectrum?  To answer this question, we turn to cosmological
simulations.  We combine the realizations of the $T_0$ and $\gamma$
fields discussed in the previous section with the density and peculiar
velocity fields from a cosmological simulation.  We then extract
absorption spectra from the simulation, and measure the flux power
with and without temperature fluctuations.

The cosmological simulation we use is a Hydro-Particle-Mesh (HPM)
simulation (\citealt{gned98}; \citealt{heit05}; \citealt{lidz05};
Habib et al.\ 2005, in prep.), with $2 \times 512^3$ particles and
$512^3$ mesh-points in an $80 \hMpc$ box.  For our purposes, we want a
simulation box that is large enough to sample the \ion{He}{3} bubble
distribution, while resolving the pressure-smoothing and thermal
broadening scales. Our simulation represents a compromise between
these requirements.  The resolution is inadequate for detailed
predictions, as shown in the convergence studies presented in the
Appendix of \citet{lidz05}. Furthermore, the accuracy of HPM has been
criticized in the literature \citep[e.g.][]{viel05b}. However, our
present goal is only to investigate how the flux power spectrum
differs with and without temperature fluctuations, rather than to make
a detailed comparison with data. For this limited purpose, we believe
our simulation is adequate. One further caveat is that the gas
pressure force, as calculated in our HPM simulation, depends on the
thermal history of the IGM. In principle, we should use the
fluctuating temperature field in our models to calculate the gas
pressure in our HPM simulation. However, we ignore this effect and use
HPM simulations calculated with a uniform temperature, including the
temperature fluctuations only when we construct the artificial \lya
forest spectra. The flux power spectrum, particularly on SDSS scales,
depends rather weakly on gas pressure-smoothing \citep{mcdo04b,
viel05c}, and so this is probably a good approximation.

We extract artificial spectra at $z = 3$ from the HPM simulation box
in the usual way (see \Eqn{Eqn::tau}), incorporating peculiar
velocities and thermal broadening.  This is done for each of our five
models, and we compare the flux power spectrum in the fluctuating
temperature model with that in a similar model without temperature
fluctuations. In the models without temperature fluctuations, $T_0$
and $\gamma$ are each set to their global averages in the
corresponding models that include temperature fluctuations.  In each
case, we normalize the quantity $A_\tau$ in \Eqn{Eqn::tau} to match
the observed mean transmitted flux at $z = 3$, which we take to be
$\avg{F} = 0.68$, close to the value measured by \citet{mcdo00}.

\begin{figure}[t]
\includegraphics[width=\columnwidth,bb=0 140 640 760]{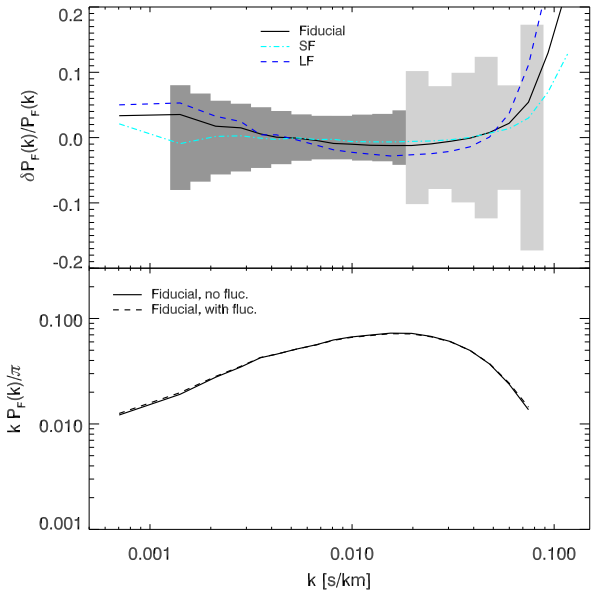}
\caption{{\it Top:} Fractional difference in the $z = 3$ flux
power spectrum between simulations with and without temperature
fluctuations.  The fiducial, SF, and LF models are plotted with the
1-$\sigma$ fractional errors from \citet{mcdo04a} and \citet{crof02}
(dark and light gray shaded areas, respectively).  {\it Bottom:} Flux
power spectrum with and without temperature fluctuations in the
fiducial model.}
\label{Fig::FPSt}
\end{figure}

\begin{figure}[t]
\includegraphics[width=\columnwidth,bb=0 140 640 760]{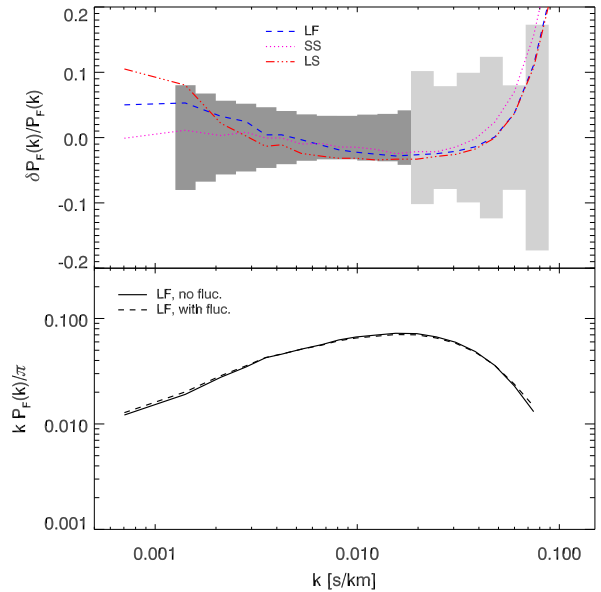}
\caption{{\it Top:} Similar to \Fig{Fig::FPSt} but for the LF,
SS, and LS models.  {\it Bottom:} Flux power spectrum with and without
temperature fluctuations in the LF model.}
\label{Fig::FPSs}
\end{figure}

The results of this calculation are shown in \Fig{Fig::FPSt} and
\Fig{Fig::FPSs}, where we parameterize the effect of temperature
fluctuations by the fractional difference in the flux power spectrum
between a fluctuating temperature model and a no fluctuations
model. The first feature to notice is simply that the effect of
temperature fluctuations on the flux power spectrum is quite small on
all scales examined. Closer examination reveals that the flux power
spectrum is boosted on large scales and on small scales compared to
equivalent models without temperature fluctuations, while there is a
very slight suppression on intermediate scales. We will discuss each
effect in turn.

The first effect is due to increased structure in the forest on large
scales, contributed by the hot, isothermal \ion{He}{3} bubbles.  The flux
power is boosted on large scales because the flux transmission is
sensitive to the temperature-dependent recombination coefficient, and
the spatially fluctuating temperature-density relation.  In
\Fig{Fig::FPSt}, concentrating on large scales for the moment, we
study the effect for models with varying levels of temperature
fluctuations: the fiducial, SF, and LF models. Specifically, we
compare the fractional difference between the flux power spectrum with
and without temperature fluctuations and the 1-$\sigma$ error bars on
the SDSS measurements of \citet{mcdo04a}. The fractional difference
between the models is always smaller than the 1-$\sigma$ SDSS
errors. The difference is only at the $5\%$ level on the largest
scales probed in the LF model. The effect also appears to diminish
rather quickly with decreasing temperature fluctuation strength, as
illustrated by the other two models in the figure.

In \Fig{Fig::FPSs}, we show the same comparison for models in which we
vary the {\em characteristic scale} of the temperature fluctuations:
the SS, LF, and LS models. Each of these models has approximately the
{\em same fluctuation strength}, and only differ in their
characteristic scale. In this case, the effect can be as large as
$\sim 10\%$ on the largest scales probed, comparable to the SDSS
1-$\sigma$ error bars at this redshift. In the models where the
characteristic scale of the temperature fluctuations is smaller, the
effect on the flux power spectrum is smaller.  This appears to
be consistent with the interpretation suggested in \Sec{Sec::TempPS}:
density fluctuations generally swamp temperature fluctuations, unless
the temperature fluctuations have a large characteristic scale.

The next effect is a boost in the small scale power, which is the
result of thermal broadening.  The figures illustrate that the models
with temperature fluctuations typically have $\sim 20\%$ more power on
scales of $k \sim 0.1$ s/km than corresponding models with a uniform
$T_0$ and $\gamma$. In \Fig{Fig::FPSt} and \ref{Fig::FPSs}, we compare
this boost in small scale power to the 1-$\sigma$ statistical
error-bars on the measurement of \citet{crof02}, which is the most
precise measurement to date on these scales.  The enhanced small scale
power in the models with temperature fluctuations can be understood as
follows. On small scales, the power spectrum in a fluctuating
temperature model will approximately be a filling-factor weighted
average of the power spectrum of hot regions and that of cold
regions. Roughly speaking, the power spectrum on small scales is
exponentially suppressed with increasing temperature
\citep{zald01b}. The weighted average we mention, and hence the power
spectrum on small scales in fluctuating temperature models, is
therefore dominated by the cold regions. The fluctuating temperature
model will then have {\em more small scale power} than a uniform
temperature model with the same {\em mean temperature} as the
fluctuating model.

Finally, there is a slight suppression in the flux power on
intermediate scales, typically around a few percent.  This suppression
occurs on scales in which the temperature field has little power, and
on scales too large for thermal broadening to have an effect. The
suppression is likely a consequence of the fact that the normalization
$A_\tau$ required to match the observed mean transmitted flux is
slightly smaller in models with a fluctuating temperature field.  The
smaller $A_\tau$ in the fluctuating temperature models implies that
density fluctuations of a given amplitude are translated into lesser
optical depth fluctuations on intermediate scales.

One might expect temperature fluctuations to have a larger effect at
high redshift, when the amplitude of density fluctuations is smaller.
We test this by considering a model that, at $z=4$, has very similar
temperature fluctuations to those in our LF model at $z=3$. The
parameters of this model, which we call LF4, are detailed in
Table~\ref{ParTab}.  We find that the effect, again parameterized by
the fractional difference between the models with and without
temperature fluctuations, is only a couple percent larger on large
scales in the LF4 model. Even though the effect is larger at $z=4$, it
is less significant in the sense that the statistical errors on the
SDSS measurement are larger at $z=4$ than at $z=3$.

\section{Conclusion} \label{conclusion}
In this paper, we have estimated the level of temperature fluctuations
expected in the IGM at $z \sim 3$ from extended \ion{H}{1} reionization and
incomplete \ion{He}{2} reionization. We find that the temperature
fluctuations from extended \ion{H}{1} reionization should be quite small,
$\sigma_{T_0}/\avg{T_0} \lesssim 5\%$, while the fluctuations from
incomplete \ion{He}{2} reionization might be as large as $\sim 50\%$.  These
fluctuations should have only a small effect on the flux power
spectrum: on large scales, $k \sim 0.001 \;{\rm s/km}$, temperature
fluctuations lead to an increase in the $z = 3$ flux power spectrum by
at most $\sim 10\%$.  On small scales, $k \sim 0.1 \;{\rm s/km}$,
fluctuations in the thermal broadening scale boost the power by $\sim
20\%$.

Further study is required to quantify the effects of temperature
fluctuations on cosmological parameter constraints. In particular,
note that in the LS model (see \Fig{Fig::FPSs}), the effect of
temperature fluctuations are at, or larger than, the 1-$\sigma$ level
over several independent data points.  A detailed investigation will
require a multi-parameter fit to the observed data
\citep[see][]{mcdo04b}, which is outside the scope of the present
paper.  However, we can anticipate the results of a more detailed
investigation by comparing with the effects of UV background
fluctuations, as studied in \citet{mcdo04c,mcdo04b}.  Fluctuations in
the UV background and temperature fluctuations have a similar
influence on the amplitude and shape of the flux power spectrum, at
least on large scales (see Fig.~12 of \citealt{mcdo04b}).
\citet{mcdo04b} included the effect of UV background fluctuations in a
multi-parameter fit, and found little effect ($\lesssim 1\%$) on the
inferred values of the amplitude and slope of the linear power
spectrum. The general reason for this insensitivity is that the
effective errors on the flux power spectrum are larger than the raw
statistical errors on the data, after \citet{mcdo04b} marginalize over
other, more significant, effects.  Similarly, we do not expect
temperature fluctuations to have a substantial effect on the values of
the amplitude and slope of the linear power spectrum.

For the purpose of cosmological parameter estimation, it would be
instructive to know whether the effect of temperature fluctuations is
degenerate with any parameters describing the \lya forest.  As we
mentioned in the previous paragraph, the effect of temperature
fluctuations may be degenerate with those of UV background
fluctuations on large scales. However, they have different effects on
small scales.  Additionally, inspecting, e.g.\ Fig.~13 of
\citet{mcdo04b} \citep[see also][]{viel05c}, shows that the boost in
the flux power on large and small scales expected from temperature
fluctuations is not closely mimicked by changing any single modeling
parameter. In principle, this means that the effect of temperature
fluctuations is likely distinguishable from other effects.  The
redshift evolution of the effect potentially provides an additional
diagnostic.  However, even though the number of quasar spectra will
likely more than double by the time SDSS is finished, detecting
temperature fluctuations in the flux power spectrum will remain a
challenge owing to the smallness of the effect.

Numerous improvements could be made in our simple modeling. First, we
place quasars at random positions in our simulation box, which ignores
quasar clustering.  This effect is already discussed in
\Sec{Sec::TempPS}.  Here, we note that since quasars are very sparse
sources, ignoring source clustering is a much better approximation
than in the case that the sources are galaxies, in which case this
approximation is quite poor \citep{furl04}.  Furthermore, in reality,
quasars should reside in very massive halos, rather than at random
positions. As a result, we ignore the fact that very close to the
quasar the gas will be overdense, tending to cancel out the enhanced
transmission owing to the hot \ion{He}{3} bubble around the quasar. The
bubbles are, however, $\sim 10\hMpc$ in size, over which the density
contrast will be small.

In modeling temperature fluctuations (the `thermal proximity effect'),
we also ignored the `radiation proximity effect'. In reality the
intensity of ionizing radiation will be enhanced over that of the
radiation background close to an {\em active} quasar
\citep[e.g.][]{scot02,roll05}. Nearby dead quasars, `light echos' of
enhanced radiation will remain, propagating out into the IGM
\citep{crof04}. These effects are ignored in our modeling, and the
radiation background is treated as uniform.  In any case, the effects
of the radiation proximity effect should be small compared to that
from temperature fluctuations.  This is because the radiation
proximity effect has a characteristic time scale of $t_q \sim 10^7$
yrs, short compared to the characteristic time scale of temperature
fluctuations, $t_{\rm cool} \sim 10^9$ yrs.  Here, we define the
cooling time $t_{\rm cool}$ to be the time it takes for a gas element
to cool to half its original temperature assuming only adiabatic
cooling.  The longer characteristic time scale, and the fact that
temperature fluctuations and the radiation proximity effect have
similar physical characteristic scales, imply that a much larger
fraction of space is affected by temperature fluctuations than by the
radiation proximity effect.

Additionally, we used a simplified `light bulb' model for quasar
activity, in which each quasar shines at constant luminosity for the
duration of its lifetime, $t_q$, and emit no light thereafter.  In
reality, this is probably a poor approximation to the quasar light
curve, since quasars likely spend extended periods at low luminosity,
going into or coming out of their peak luminosity phase
\citep{hopk05a, spri05}. Finally, quasars may launch large outflows
\citep[e.g.][]{scan04,dim05}, which could also modify the absorption in
their vicinity.

In spite of all of these possible complications, we strove to cover a
wide range in the amplitude and scale of temperature fluctuations in
our analysis.  It is unlikely that the problems discussed above will
lead to an effect that lies far outside the range probed by our study.
It should therefore be a fairly secure conclusion that temperature
fluctuations do not significantly impact the \lya forest flux power
spectrum on large scales. Our results provide additional support for
the \lya forest as a robust probe of cosmology.

Even though temperature fluctuations seem to have a small effect on
the flux power spectrum, they may have a larger effect on other
observables of the \lya forest.  The flux power spectrum, while being
the best measured statistic in the \lya forest, may be a poor
statistic to use in searching for temperature fluctuations. There have
been several searches for temperature fluctuations in the \lya forest
with wavelet analyses \citep{theu02b, zald02}.  These searches have
failed to detect temperature fluctuations, but have been carried out
using only very small data samples.  It would be interesting to
investigate the expected signal from these searches given our
modeling. A related statistic that might be sensitive to temperature
fluctuations is the \lya forest `tri-spectrum', as defined in
\citet{zald01a} (see also \citealt{fang04}). This statistic measures
the scatter in the small scale power, as a function of scale, from
region to region in the IGM.  We might, therefore, expect a larger
tri-spectrum in our models with temperature fluctuations than in
models with no temperature fluctuations.

One final point is that if \ion{He}{2} reionization is underway at $z = 3$,
this might cause biases in estimates of the IGM temperature from the
flux power spectrum, and the ionizing background derived from the \lya
forest proximity effect.  The enhancement we find in the flux power
spectrum on small scales might imply a slight bias in measurements of
the IGM temperature from the small scale flux power spectrum
\citep[e.g.][]{zald01b}, and other similar measurements. A crude
estimate of the bias is as follows.  Approximately, thermal broadening
suppresses the flux power exponentially so that $P_F(k) \propto
\exp(-k^2\sigma^2)$, where $\sigma^2 = k_{\rm b} T/m_{\rm H}$, and
$k_{\rm b}$ is Boltzmann's constant. We found that temperature
fluctuations increase $P_F(k)$ at $k = 0.1$ s/km by $20\%$, which
thereby implies a $\sim 10\%$ {\em underestimate} of the temperature
in models that assume a uniform temperature. We caution, however, that
the mean temperature is an incomplete description of the IGM thermal
state in the presence of inhomogeneous reionization and/or large
quantities of hot, shocked gas.  Temperature fluctuations may also
bias estimates of the ionizing background from the quasar proximity
effect, which assumes that the IGM is at the cosmic mean temperature
close to the quasar. We are investigating this, and other possible
biases in the constraints from the proximity effect, using radiative
transfer simulations.

\phantom{}

AL thanks Katrin Heitmann and Salman Habib for their collaboration in
producing the HPM simulation used in this analysis. We thank the
anonymous referee for useful comments on our manuscript. We also thank
Scott Burles, Steve Furlanetto, John Huchra, and Peng Oh for useful
discussions on these, and related topics.  This work was supported in
part by NSF grants ACI 96-19019, AST 00-71019, AST 02-06299, AST
03-07690, and NASA ATP grants NAG5-12140, NAG5-13292, and NAG5-13381.
Some of the simulations were performed at the Center for Parallel
Astrophysical Computing at the Harvard-Smithsonian Center for
Astrophysics.

\bibliographystyle{apj}
\bibliography{ms}

\begin{thebibliography}{68}
\expandafter\ifx\csname natexlab\endcsname\relax\def\natexlab#1{#1}\fi

\bibitem[{{Abel} \& {Haehnelt}(1999)}]{abel99}
{Abel}, T. \& {Haehnelt}, M.~G. 1999, \apjl, 520, L13

\bibitem[{{Babich} \& {Loeb}(2006)}]{babi05}
{Babich}, D. \& {Loeb}, A. 2006, \apj, 640, 1

\bibitem[{{Barkana} \& {Loeb}(2004)}]{bark04}
{Barkana}, R. \& {Loeb}, A. 2004, \apj, 609, 474

\bibitem[{{Bi} \& {Davidsen}(1997)}]{bi97}
{Bi}, H. \& {Davidsen}, A.~F. 1997, \apj, 479, 523

\bibitem[{{Bond} {et~al.}(1991){Bond}, {Cole}, {Efstathiou}, \&
  {Kaiser}}]{bond91}
{Bond}, J.~R., {Cole}, S., {Efstathiou}, G., \& {Kaiser}, N. 1991, \apj, 379,
  440

\bibitem[{{Bond} \& {Wadsley}(1997)}]{bond97}
{Bond}, J.~R. \& {Wadsley}, J.~W. 1997, in ASP Conf. Ser. 123: Computational
  Astrophysics; 12th Kingston Meeting on Theoretical Astrophysics, 323--+

\bibitem[{{Boyle} {et~al.}(1988){Boyle}, {Shanks}, \& {Peterson}}]{boyl98}
{Boyle}, B.~J., {Shanks}, T., \& {Peterson}, B.~A. 1988, \mnras, 235, 935

\bibitem[{{Bryan} {et~al.}(1999){Bryan}, {Machacek}, {Anninos}, \&
  {Norman}}]{brya99}
{Bryan}, G.~L., {Machacek}, M., {Anninos}, P., \& {Norman}, M.~L. 1999, \apj,
  517, 13

\bibitem[{{Cen} {et~al.}(1994){Cen}, {Miralda-Escude}, {Ostriker}, \&
  {Rauch}}]{cen94}
{Cen}, R., {Miralda-Escude}, J., {Ostriker}, J.~P., \& {Rauch}, M. 1994, \apjl,
  437, L9

\bibitem[{{Croft}(2004)}]{crof04}
{Croft}, R.~A.~C. 2004, \apj, 610, 642

\bibitem[{{Croft} {et~al.}(1999){Croft}, {Hu}, \& {Dav{\' e}}}]{crof99}
{Croft}, R.~A.~C., {Hu}, W., \& {Dav{\' e}}, R. 1999, Physical Review Letters,
  83, 1092

\bibitem[{{Croft} {et~al.}(2002){Croft}, {Weinberg}, {Bolte}, {Burles},
  {Hernquist}, {Katz}, {Kirkman}, \& {Tytler}}]{crof02}
{Croft}, R.~A.~C., {Weinberg}, D.~H., {Bolte}, M., {Burles}, S., {Hernquist},
  L., {Katz}, N., {Kirkman}, D., \& {Tytler}, D. 2002, \apj, 581, 20

\bibitem[{{Croft} {et~al.}(1998){Croft}, {Weinberg}, {Katz}, \&
  {Hernquist}}]{crof98}
{Croft}, R.~A.~C., {Weinberg}, D.~H., {Katz}, N., \& {Hernquist}, L. 1998,
  \apj, 495, 44

\bibitem[{{Croom} {et~al.}(2005){Croom}, {Boyle}, {Shanks}, {Smith}, {Miller},
  {Outram}, {Loaring}, {Hoyle}, \& {da {\^ A}ngela}}]{croo05}
{Croom}, S.~M., {et~al.} 2005,
  \mnras, 356, 415

\bibitem[{{Croom} {et~al.}(2004){Croom}, {Smith}, {Boyle}, {Shanks}, {Miller},
  {Outram}, \& {Loaring}}]{croo04}
{Croom}, S.~M., {Smith}, R.~J., {Boyle}, B.~J., {Shanks}, T., {Miller}, L.,
  {Outram}, P.~J., \& {Loaring}, N.~S. 2004, \mnras, 349, 1397

\bibitem[{{Dav{\'e}} {et~al.}(1999){Dav{\'e}}, {Hernquist}, {Katz}, \&
  {Weinberg}}]{dave99}
{Dav{\'e}}, R., {Hernquist}, L., {Katz}, N., \& {Weinberg}, D.~H. 1999, \apj,
  511, 521

\bibitem[{{Di Matteo} {et~al.}(2005){Di Matteo}, {Springel}, \&
  {Hernquist}}]{dim05}
{Di Matteo}, T., {Springel}, V., \& {Hernquist}, L. 2005, \nat, 433, 604

\bibitem[{{Fan} {et~al.}(2002){Fan}, {Narayanan}, {Strauss}, {White}, {Becker},
  {Pentericci}, \& {Rix}}]{fan02}
{Fan}, X., {Narayanan}, V.~K., {Strauss}, M.~A., {White}, R.~L., {Becker},
  R.~H., {Pentericci}, L., \& {Rix}, H.-W. 2002, \aj, 123, 1247

\bibitem[{{Fan} {et~al.}(2001){Fan}, {Strauss}, {Schneider}, {Gunn}, {Lupton},
  {Becker}, {Davis}, {Newman}, {Richards}, {White}, {Anderson}, {Annis},
  {Bahcall}, {Brunner}, {Csabai}, {Hennessy}, {Hindsley}, {Fukugita}, {Kunszt},
  {Ivezi{\' c}}, {Knapp}, {McKay}, {Munn}, {Pier}, {Szalay}, \&
  {York}}]{fan01a}
{Fan}, X., {et~al.} 2001, \aj, 121, 54

\bibitem[{{Fang} \& {White}(2004)}]{fang04}
{Fang}, T. \& {White}, M. 2004, \apjl, 606, L9

\bibitem[{{Furlanetto} {et~al.}(2006){Furlanetto}, {McQuinn}, \&
  {Hernquist}}]{furl05a}
{Furlanetto}, S.~R., {McQuinn}, M., \& {Hernquist}, L. 2006, \mnras, 365, 115

\bibitem[{{Furlanetto} \& {Oh}(2005)}]{furl05b}
{Furlanetto}, S.~R. \& {Oh}, S.~P. 2005, \mnras, 363, 1031

\bibitem[{{Furlanetto} {et~al.}(2004){Furlanetto}, {Zaldarriaga}, \&
  {Hernquist}}]{furl04}
{Furlanetto}, S.~R., {Zaldarriaga}, M., \& {Hernquist}, L. 2004, \apj, 613, 1

\bibitem[{{Gnedin} \& {Hui}(1998)}]{gned98}
{Gnedin}, N.~Y. \& {Hui}, L. 1998, \mnras, 296, 44

\bibitem[{{Heitmann} {et~al.}(2005){Heitmann}, {Ricker}, {Warren}, \&
  {Habib}}]{heit05}
{Heitmann}, K., {Ricker}, P.~M., {Warren}, M.~S., \& {Habib}, S. 2005, \apjs,
  160, 28

\bibitem[{{Hernquist} {et~al.}(1996){Hernquist}, {Katz}, {Weinberg}, \&
  {Miralda-Escud{\' e}}}]{hern96}
{Hernquist}, L., {Katz}, N., {Weinberg}, D.~H., \& {Miralda-Escud{\' e}}, J.
  1996, \apjl, 457, L51+

\bibitem[{{Hopkins} {et~al.}(2005){Hopkins}, {Hernquist}, {Cox}, {Di Matteo},
  {Robertson}, \& {Springel}}]{hopk05a}
{Hopkins}, P.~F., {Hernquist}, L., {Cox}, T.~J., {Di Matteo}, T., {Robertson},
  B., \& {Springel}, V. 2005, \apj, 630, 716

\bibitem[{{Hui} \& {Gnedin}(1997)}]{hui97b}
{Hui}, L. \& {Gnedin}, N.~Y. 1997, \mnras, 292, 27

\bibitem[{{Hui} {et~al.}(1997){Hui}, {Gnedin}, \& {Zhang}}]{hui97a}
{Hui}, L., {Gnedin}, N.~Y., \& {Zhang}, Y. 1997, \apj, 486, 599

\bibitem[{{Hui} \& {Haiman}(2003)}]{hui03}
{Hui}, L. \& {Haiman}, Z. 2003, \apj, 596, 9

\bibitem[{{Lacey} \& {Cole}(1993)}]{lace93}
{Lacey}, C. \& {Cole}, S. 1993, \mnras, 262, 627

\bibitem[{{Lidz} {et~al.}(2006){Lidz}, {Heitmann}, {Hui}, {Habib}, {Rauch}, \&
  {Sargent}}]{lidz05}
{Lidz}, A., {Heitmann}, K., {Hui}, L., {Habib}, S., {Rauch}, M., \& {Sargent},
  W.~L.~W. 2006, \apj, 638, 27

\bibitem[{{Madau} {et~al.}(1999){Madau}, {Haardt}, \& {Rees}}]{mada99}
{Madau}, P., {Haardt}, F., \& {Rees}, M.~J. 1999, \apj, 514, 648

\bibitem[{{McDonald} {et~al.}(2000){McDonald}, {Miralda-Escud{\' e}}, {Rauch},
  {Sargent}, {Barlow}, {Cen}, \& {Ostriker}}]{mcdo00}
{McDonald}, P., {Miralda-Escud{\' e}}, J., {Rauch}, M., {Sargent}, W.~L.~W.,
  {Barlow}, T.~A., {Cen}, R., \& {Ostriker}, J.~P. 2000, \apj, 543, 1

\bibitem[{{McDonald} {et~al.}(2005{\natexlab{a}}){McDonald}, {Seljak}, {Cen},
  {Bode}, \& {Ostriker}}]{mcdo04c}
{McDonald}, P., {Seljak}, U., {Cen}, R., {Bode}, P., \& {Ostriker}, J.~P.
  2005{\natexlab{a}}, \mnras, 360, 1471

\bibitem[{{McDonald} {et~al.}(2005{\natexlab{b}}){McDonald}, {Seljak}, {Cen},
  {Shih}, {Weinberg}, {Burles}, {Schneider}, {Schlegel}, {Bahcall}, {Briggs},
  {Brinkmann}, {Fukugita}, {Ivezi{\'c}}, {Kent}, \& {Vanden Berk}}]{mcdo04b}
{McDonald}, P., {et~al.} 2005{\natexlab{b}}, \apj, 635, 761

\bibitem[{{McDonald} {et~al.}(2006){McDonald}, {Seljak}, {Burles}, {Schlegel},
  {Weinberg}, {Shih}, {Schaye}, {Schneider}, {Brinkmann}, {Brunner}, \&
  {Fukugita}}]{mcdo04a}
{McDonald}, P., {et~al.} 2006, \apjs, 163, 80

\bibitem[{{Meiksin}(2005)}]{meik05}
{Meiksin}, A. 2005, \mnras, 356, 596

\bibitem[{{Meiksin} \& {White}(2004)}]{meik04}
{Meiksin}, A. \& {White}, M. 2004, \mnras, 350, 1107

\bibitem[{{Miralda-Escude} {et~al.}(1996){Miralda-Escude}, {Cen}, {Ostriker},
  \& {Rauch}}]{mira96}
{Miralda-Escude}, J., {Cen}, R., {Ostriker}, J.~P., \& {Rauch}, M. 1996, \apj,
  471, 582

\bibitem[{{Miralda-Escude} \& {Rees}(1994)}]{mira94}
{Miralda-Escude}, J. \& {Rees}, M.~J. 1994, \mnras, 266, 343

\bibitem[{{Muecket} {et~al.}(1996){Muecket}, {Petitjean}, {Kates}, \&
  {Riediger}}]{muec96}
{Muecket}, J.~P., {Petitjean}, P., {Kates}, R.~E., \& {Riediger}, R. 1996,
  \aap, 308, 17

\bibitem[{{Nusser} \& {Haehnelt}(1999)}]{nuss99}
{Nusser}, A. \& {Haehnelt}, M. 1999, \mnras, 303, 179

\bibitem[{{Peacock} \& {Dodds}(1996)}]{peac96}
{Peacock}, J.~A. \& {Dodds}, S.~J. 1996, \mnras, 280, L19

\bibitem[{{Pei}(1995)}]{pei95}
{Pei}, Y.~C. 1995, \apj, 438, 623

\bibitem[{{Richards} {et~al.}(2005){Richards}, {Croom}, {Anderson},
  {Bland-Hawthorn}, {Boyle}, {De Propris}, {Drinkwater}, {Fan}, {Gunn},
  {Ivezi{\' c}}, {Jester}, {Loveday}, {Meiksin}, {Miller}, {Myers}, {Nichol},
  {Outram}, {Pimbblet}, {Roseboom}, {Ross}, {Schneider}, {Shanks}, {Sharp},
  {Stoughton}, {Strauss}, {Szalay}, {Vanden Berk}, \& {York}}]{rich05}
{Richards}, G.~T., {et~al.} 2005, \mnras, 360, 839

\bibitem[{{Rollinde} {et~al.}(2005){Rollinde}, {Srianand}, {Theuns},
  {Petitjean}, \& {Chand}}]{roll05}
{Rollinde}, E., {Srianand}, R., {Theuns}, T., {Petitjean}, P., \& {Chand}, H.
  2005, \mnras, 610

\bibitem[{{Scannapieco} \& {Oh}(2004)}]{scan04}
{Scannapieco}, E. \& {Oh}, S.~P. 2004, \apj, 608, 62

\bibitem[{{Scott} {et~al.}(2002){Scott}, {Bechtold}, {Morita}, {Dobrzycki}, \&
  {Kulkarni}}]{scot02}
{Scott}, J., {Bechtold}, J., {Morita}, M., {Dobrzycki}, A., \& {Kulkarni},
  V.~P. 2002, \apj, 571, 665

\bibitem[{{Seljak} {et~al.}(2005){Seljak}, {Makarov}, {McDonald}, {Anderson},
  {Bahcall}, {Brinkmann}, {Burles}, {Cen}, {Doi}, {Gunn}, {Ivezi{\' c}},
  {Kent}, {Loveday}, {Lupton}, {Munn}, {Nichol}, {Ostriker}, {Schlegel},
  {Schneider}, {Tegmark}, {Berk}, {Weinberg}, \& {York}}]{selj05}
{Seljak}, U., {et~al.} 2005, \prd, 71, 103515

\bibitem[{{Shapiro} \& {Giroux}(1987)}]{shap87}
{Shapiro}, P.~R. \& {Giroux}, M.~L. 1987, \apjl, 321, L107

\bibitem[{{Sokasian} {et~al.}(2002){Sokasian}, {Abel}, \& {Hernquist}}]{soka02}
{Sokasian}, A., {Abel}, T., \& {Hernquist}, L. 2002, \mnras, 332, 601

\bibitem[{{Sokasian} {et~al.}(2003){Sokasian}, {Abel}, {Hernquist}, \&
  {Springel}}]{soka03}
{Sokasian}, A., {Abel}, T., {Hernquist}, L., \& {Springel}, V. 2003, \mnras,
  344, 607

\bibitem[{{Sokasian} {et~al.}(2004){Sokasian}, {Yoshida}, {Abel}, {Hernquist},
  \& {Springel}}]{soka04}
{Sokasian}, A., {Yoshida}, N., {Abel}, T., {Hernquist}, L., \& {Springel}, V.
  2004, \mnras, 350, 47

\bibitem[{{Springel} {et~al.}(2005){Springel}, {Di Matteo}, \&
  {Hernquist}}]{spri05}
{Springel}, V., {Di Matteo}, T., \& {Hernquist}, L. 2005, \mnras, 361, 776

\bibitem[{{Theuns} {et~al.}(1999){Theuns}, {Leonard}, {Schaye}, \&
  {Efstathiou}}]{theu99}
{Theuns}, T., {Leonard}, A., {Schaye}, J., \& {Efstathiou}, G. 1999, \mnras,
  303, L58

\bibitem[{{Theuns} {et~al.}(2002{\natexlab{a}}){Theuns}, {Schaye}, {Zaroubi},
  {Kim}, {Tzanavaris}, \& {Carswell}}]{theu02a}
{Theuns}, T., {Schaye}, J., {Zaroubi}, S., {Kim}, T., {Tzanavaris}, P., \&
  {Carswell}, B. 2002{\natexlab{a}}, \apjl, 567, L103

\bibitem[{{Theuns} {et~al.}(2002{\natexlab{b}}){Theuns}, {Zaroubi}, {Kim},
  {Tzanavaris}, \& {Carswell}}]{theu02b}
{Theuns}, T., {Zaroubi}, S., {Kim}, T., {Tzanavaris}, P., \& {Carswell}, R.~F.
  2002{\natexlab{b}}, \mnras, 332, 367

\bibitem[{{Tytler} {et~al.}(2004){Tytler}, {Kirkman}, {O'Meara}, {Suzuki},
  {Orin}, {Lubin}, {Paschos}, {Jena}, {Lin}, {Norman}, \& {Meiksin}}]{tytl04}
{Tytler}, D., {et~al.} 2004, \apj, 617, 1

\bibitem[{{Viel} \& {Haehnelt}(2006)}]{viel05c}
{Viel}, M. \& {Haehnelt}, M.~G. 2006, \mnras, 365, 231

\bibitem[{{Viel} {et~al.}(2004){Viel}, {Haehnelt}, \& {Springel}}]{viel04c}
{Viel}, M., {Haehnelt}, M.~G., \& {Springel}, V. 2004, \mnras, 354, 684

\bibitem[{{Viel} {et~al.}(2005{\natexlab{a}}){Viel}, {Haehnelt}, \&
  {Springel}}]{viel05b}
---. 2006, \mnras, 367, 1655

\bibitem[{{Viel} {et~al.}(2005{\natexlab{b}}){Viel}, {Lesgourgues}, {Haehnelt},
  {Matarrese}, \& {Riotto}}]{viel05a}
{Viel}, M., {Lesgourgues}, J., {Haehnelt}, M.~G., {Matarrese}, S., \& {Riotto},
  A. 2005{\natexlab{b}}, \prd, 71, 063534

\bibitem[{{Zaldarriaga}(2002)}]{zald02}
{Zaldarriaga}, M. 2002, \apj, 564, 153

\bibitem[{{Zaldarriaga} {et~al.}(2001{\natexlab{a}}){Zaldarriaga}, {Hui}, \&
  {Tegmark}}]{zald01b}
{Zaldarriaga}, M., {Hui}, L., \& {Tegmark}, M. 2001{\natexlab{a}}, \apj, 557,
  519

\bibitem[{{Zaldarriaga} {et~al.}(2003){Zaldarriaga}, {Scoccimarro}, \&
  {Hui}}]{zald03}
{Zaldarriaga}, M., {Scoccimarro}, R., \& {Hui}, L. 2003, \apj, 590, 1

\bibitem[{{Zaldarriaga} {et~al.}(2001{\natexlab{b}}){Zaldarriaga}, {Seljak}, \&
  {Hui}}]{zald01a}
{Zaldarriaga}, M., {Seljak}, U., \& {Hui}, L. 2001{\natexlab{b}}, \apj, 551, 48

\bibitem[{{Zhang} {et~al.}(1995){Zhang}, {Anninos}, \& {Norman}}]{zhan95}
{Zhang}, Y., {Anninos}, P., \& {Norman}, M.~L. 1995, \apjl, 453, L57+

\end{thebibliography}

\end{document}